\documentclass[conference]{IEEEtran}
\usepackage{hyperref}
\usepackage{optidef}
\usepackage{multirow}
\usepackage{xcolor,soul}
\usepackage[utf8]{inputenc}
\usepackage[T1]{fontenc}
\usepackage{algorithm}
\usepackage{algpseudocode}

\ifCLASSINFOpdf
  \usepackage[pdftex]{graphicx}
  \graphicspath{{./assets/}}
\else
  \graphicspath{./assets/}
\fi

\begin{document}

\title{Sizing and Profitability of Energy Storage for Prosumers in Madeira, Portugal}

 \author{\IEEEauthorblockN{Md Umar Hashmi\IEEEauthorrefmark{1},
Jonathan Cavaleiro\IEEEauthorrefmark{2},
 Lucas Pereira\IEEEauthorrefmark{3}, and
 Ana Bu\v{s}i\'c\IEEEauthorrefmark{1}}
 \IEEEauthorblockA{\IEEEauthorrefmark{1} INRIA and the Computer Science Dept. of Ecole Normale Sup\'erieure, CNRS, PSL University,
 Paris, France}
 \IEEEauthorblockA{\IEEEauthorrefmark{2} ITI, LARSyS, M-ITI Funchal, Madeira}
 \IEEEauthorblockA{\IEEEauthorrefmark{3} ITI, LARSyS, Técnico Lisboa and prsma.com, Funchal, Portugal}
 }

\maketitle

\begin{abstract}
This paper proposes a framework to select the best-suited battery for co-optimizing for peak demand shaving, energy arbitrage and increase self-sufficiency in the context of power network in Madeira, Portugal. Feed-in-tariff for electricity network in Madeira is zero, which implies consumers with excess production should locally consume the excess generation rather than wasting it. Further, the power network {operator} applies a peak power contract for consumers which imposes an upper bound on the peak power seen by the power grid interfaced by energy meter. We investigate the value of storage in Madeira, using four different types of prosumers, categorized based on the relationship between their inelastic load and renewable generation.
We observe that the marginal increase in the value of storage deteriorates with increase in size and ramping capabilities.
We propose the use of profit per cycle per unit of battery capacity and expected payback period as indices for selecting the best-suited storage parameters to ensure profitability. This mechanism takes into account the consumption and generation patterns, profit, storage degradation, and cycle and calendar life of the battery.
We also propose the inclusion of a friction coefficient in the original co-optimization formulation to increase the value of storage by reducing the operational cycles and eliminate low returning transactions.
\end{abstract}
\section{Introduction}
Renewable integration will increase opportunities for consumers in power networks for reducing their local cost of consumption \cite{hashmi2018effect}.
Energy storage co-optimization is becoming popular with the possibility of doing more than one dedicated task without undermining the profit of storage
by performing each task at a time. Authors in \cite{shi2018using} show that co-optimizing profit leads to higher than doing one or few of tasks together.

In our prior work \cite{hashmi2019energy} we introduced power network rules in Madeira and show the profitability of a battery for a local prosumer. In this paper, we explore the dependencies of inelastic load and renewable generation size on the storage parameters and its profitability.
Energy storage performing dedicated roles may not be profitable. Authors in \cite{Kied1910:Sensitivity}, \cite{hashmi2017optimal} show that batteries are not profitable if only used for energy arbitrage. Authors in \cite{Hash1910:Optimal} observe that the profit made by storage performing only arbitrage is sensitive to the ratio of selling and buying price of electricity. It is observed that as this ratio goes to zero, the value of storage increase for cases where the prosumer has inelastic consumption.
Thus in the context of Madeira energy storage can help in increasing installed renewable generation as excess generation is not compensated and in this case, a battery can be used to locally increase self-sufficiency for the prosumer.
It is noteworthy that the co-optimization of energy storage usage has been extensively researched due to the high cost of batteries. In Table~\ref{tabroles} we list some of the works in this area. In this paper, we use energy storage for performing energy arbitrage, increase self-sufficiency and facilitate peak demand shaving.


%

Authors in \cite{oudalov2007sizing} observe that peak demand shaving for discharge period of more than an hour is not profitable. 
For Madeira the Peak Power Contracts (PPCs) are applied for the instantaneous power level in the time-scale of seconds, thus use of storage for peak shaving is reasonable.
Authors in \cite{shen2014optimization} propose battery cycle life optimization for prolonging the operational life of the storage. 
Authors in \cite{bahramirad2012reliability, chen2011sizing} 
propose sizing of energy storage in context of a microgrid using 
mixed-integer programming.
\cite{bahramirad2012reliability} consider microgrid reliability for sizing and \cite{chen2011sizing} use storage devices for spinning reserves. 
Authors in \cite{brekken2010optimal} consider storage sizing in the context of wind power applications. In this paper, we provide a framework for identifying storage profitability battery for prosumers.
\begin{table}[!htbp]
	\scriptsize
	\caption {\small{Co-optimization of energy storage}}
	\label{tabroles}
	\vspace{-15pt}
	\begin{center}
		\begin{tabular}{| c | c| }
			\hline
			Paper& Co-optimizing for  \\ 
			\hline
			\cite{megel2014scheduling}& PV integration support, peak shaving, frequency regulation\\
			\hline
			\cite{walawalkar2007economics}, \cite{anderson2017co, cheng2016co}& Arbitrage and frequency regulation\\
			\hline
			\cite{shi2018using, white2011using} & Peak shaving and frequency regulation\\
			\hline
			\cite{dvorkin2016ensuring} & Arbitrage, congestion relief and curtailment reduction\\
			\hline
			\cite{hashmi2019arbitrage} & Arbitrage with power factor correction.\\
			\hline
		\end{tabular}
		\hfill\
	\end{center}
\end{table}

Li-Ion battery life is measured in cycle and calendar life. In \cite{hashmi2018limiting} the operational life of a battery is increased by matching calendar life degradation with the ageing degradation of the battery. Extending this idea and taking the depth-of-discharge (DoD) of the battery authors in \cite{hashmi2018long} propose a way to measure the cycles of operation of the battery. It is highlighted that the nonlinear relationship between cycle life and DoD, \cite{mackley2015technical,shen2014optimization}, encourages us to use this methodology to calculate storage returns per cycle. In this paper, we improve the factor \textit{profit per cycles} proposed in \cite{hashmi2018long} to make it independent of storage capacity. 
This new factor along with expected payback period of the battery can be used for identifying the expected break-even profit per cycle, thus by comparing these two scalars we can comment on 
the profitability of a battery.

Further, battery prices are falling with more research in this domain, which will promote increased usage of batteries in power networks.
Historically, the cost of battery fell by 85\% from 2010 to 2018, reaching an average price of \$176/kWh.
The price of an average Li-Ion battery pack is projected to be around \$94/kWh by 2024 and \$62/kWh by 2030 \cite{bnefbattery}.

The key contributions of this paper are:\\
$\quad \bullet$ \textit{Index for measuring the profitability of Li-Ion battery}: 
Taking into account storage parameters and profit, we propose using \textit{profit per cycle per unit capacity} and expected payback period as the factors for selecting storage parameters ensuring profitability. These factors take into account battery degradation, profit, cycle and calendar life of the battery. 
For calculating operational cycles we use algorithm \texttt{DoDofVector} proposed in \cite{hashmi2018long}.     \\
$\quad \bullet$ \textit{Numerical evaluation}: Based on consumer load and renewable generation consumer are categorized into four categories.
Using these four categories of prosumers we use the data collected for June 2019 and apply time-of-use (ToU) rates and Peak Power Contract (PPC) improvement to identify the marginal value of storage. We observe that a smaller sized and ramping capability storage is profitable in all four categories.\\
$\quad \bullet$ \textit{Observations}: Using data and numerical results we make the following observations which can be generally of interest: \vspace{-2pt}
\begin{itemize}
	\item The marginal value of storage reduces with increase in size and ramping capability. 
	\item Under zero-feed-in tariff, a higher average net-load significantly improves the profit prosumer can make.
\end{itemize}

This paper is organized as follows.
In Section II we introduce the Madeira data collection initiative under H2020 project SMILE.
In Section III we provide the system description, battery model and we summarize the co-optimization formulation proposed in \cite{hashmi2019energy} and extend this formulation for controlling cycles of operation for increasing the operational life of the battery.
In Section IV we present an algorithm for identifying the profitability of a battery.
In Section V we present the numerical results. Section VI concludes the paper.




\begin{figure*}[!htbp]
	\centering
	\includegraphics[width=4.7in]{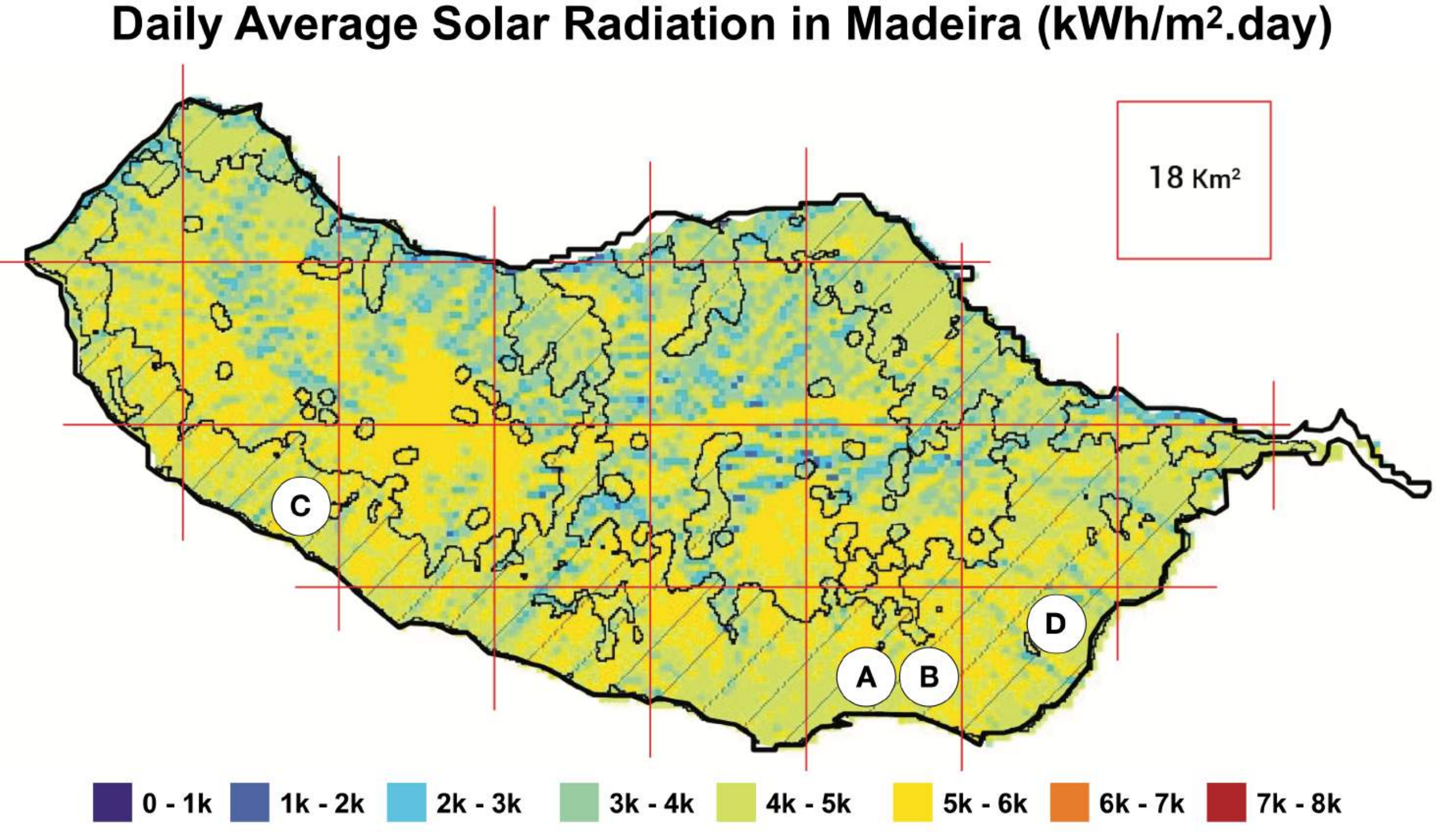}
	\vspace{-14pt}
	\caption{Location of the four prosumers in Madeira used in this work}
	\label{fig:upaclocation}
\end{figure*}

\section{Context and Data Collection in Madeira}

%

In the Madeira archipelago, since 2014, prosumers are not allowed to inject the excess energy to the local grid. This imposition is owing to the isolated nature of the Madeira electric grid that is very sensitive to variations in the energy produced by RES. 
Yet, while this technical restriction helps to maintain grid stability, it is also leading to a stagnation in the number of prosumers. For example, as of this writing, there are only 49 solar PV prosumers registered in Madeira island (an average of about 0.6 kWp installed capacity), in a universe of 270,000 residents \cite{hashmi2019energy,pereira_value_2019}.
{Inclusion of energy storage along with over-sizing renewables at consumer end have the potential to increase renewable installations in the island.}

The development of better control strategies for battery energy storage is one of the goals of the H2020 SMILE, \url{https://www.h2020smile.eu}, Project, an European Union co-funded project.
Under the scope of this project, PV production ($PV$) and power consumption ($Load$) measurements were taken from 14 prosumers in the island \cite{prsma_data_2018,prsma_detailed_2018}. 
The monitored prosumers are categorized into four categories, based on their inelastic load and PV generation: (C1) PV generation slightly more than the inelastic load leading to some waste of excess generation, (C2) consumer actively managing the load to match the instantaneous PV generation thus reducing the production wastage, (C3) PV generation is comparable to the magnitude of inelastic load with significantly higher average net-load compared to other cases considered and (C4) PV generation is significantly more than the inelastic load leading to significant amount of energy wastage.
The location of these prosumers are shown on the map of Madeira in Fig.~\ref{fig:upaclocation}.

For this paper, one prosumer was selected from each category. The PPC, installed solar PV capacity, and the monthly totals of each prosumer for June 2019 are presented in Table \ref{tab:upac_details}. 

In Table~\ref{tab:upac_details}, $PV_{Loads}$ is the amount of solar PV production that is consumed directly by the loads, $PV_{Grid}$ is the amount of solar PV production injected into the grid (hereby considered wasted production since there is no feed-in tariff), $Grid_{Loads}$ is the amount of power from the grid that is consumed by the loads, and $Savings$ is the amount of saving due to the installation of the solar PV (assuming a fixed cost of 0.16 Euro per kWh).

\begin{figure}[!htbp]
	\center
	\includegraphics[width=3.4in]{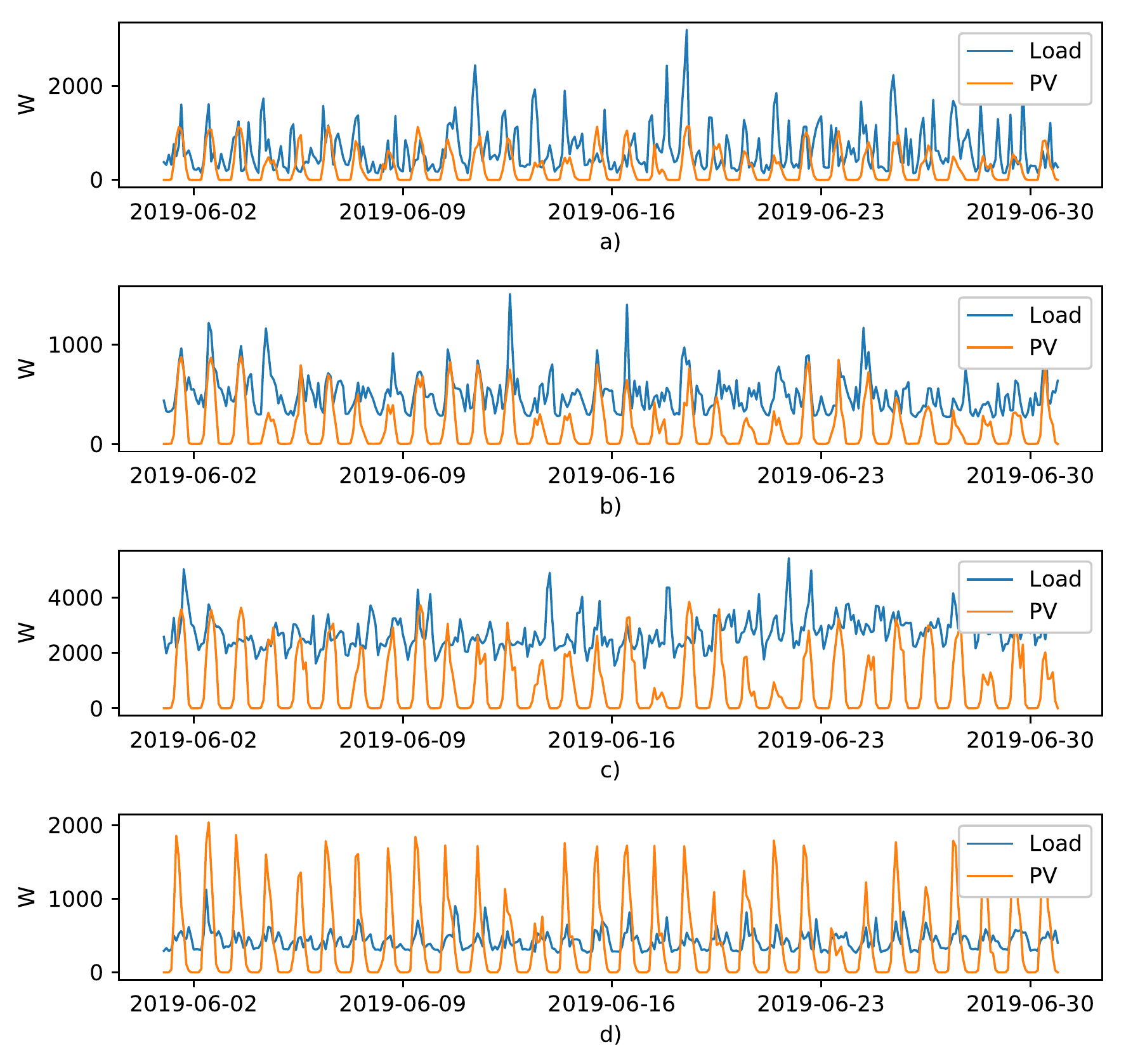}
	\vspace{-14pt}
	\caption{PV and inelastic load for 4 types of prosumers C1, C2, C3 and C4}
	\label{fig:upac_pvload}
\end{figure}

Fig.~\ref{fig:upac_pvload} shows the relationship between PV and inelastic load for the selected prosumers. U2, U3, and U12 are domestic prosumers. U8 is a commercial prosumer, a small family restaurant.

\begin{table}[ht]
\scriptsize
\centering
\caption {Indices of load and generation for U2, U3, U8, U12 for June 2019}
\begin{tabular}{|c|c|c|c|c|c|c|}
\hline
Case &UPAC & Contracted & Installed & PV & Load & $PV_{Load}$  \\
&ID& Power (kVA) & PV (kWp) & (kWh) & (kWh) & (kWh)  \\
\hline
C1 & U2  & 6.9 & 1.5 & 180 & 459 & 121  \\\hline
C2 &U3  & 5.75 & 1 & 120 & 360 & 113 \\\hline
C3 & U8  & 20.7 & 3.92 & 610 & 1969 & 531\\\hline
C4 & U12 & 6.9 & 3 & 305 & 306 & 145  \\\hline
\hline
Case &UPAC &  $PV_{Grid}$ & $Grid_{Load}$ & Savings  & Average & Average  \\
&ID& (kWh) & (kWh) & (Euros) & PV (W) & Load (W)  \\
\hline
C1 & U2  & 59.64 & 338 & 19.30 & 251 & 637 \\\hline
C2 &U3  &  6.85 & 247 & 21.19 & 168 & 501 \\\hline
C3 & U8  &  78.53 & 1438 & 85.09 & 848 & 2736  \\\hline
C4 & U12 &  160 & 162 & 26.37 & 424 & 426 \\\hline
\end{tabular}
\label{tab:upac_details}
\end{table}
%

\section{System Model and Storage Co-Optimization}
In this section, we present the system and battery model and the summarize the co-optimization formulation developed in \cite{hashmi2019energy}. We extend this formulation for limiting storage cycles of operation using a friction coefficient.


\subsection{System and Battery Model}
We consider operation over a total duration $T$, divided into $N$ steps indexed by $\{1,...,N\}$. The duration of each step is denoted as $h$. Hence, $T=hN$. 
At time instant $i$, the information available is the end user consumption $d_i$, the renewable generation ${r}_i$ and the storage energy output $s_i$. 
The load without storage is denoted as
$
z_i = d_i - r_i.
$
The load seen by grid is denoted as
$
L_i = d_i - r_i + s_i.
$

\textit{Battery Model:} The efficiency of charging and discharging 
is denoted by $\eta_{ch}, \eta_{dis} \in (0,1]$, respectively.
The change in energy level of the battery is denoted as 
$x_i$=$h \delta_i$, where $\delta_i$ denotes storage ramp rate at time instant $i$,
%
such that $\delta_i \in [\delta_{\min}, \delta_{\max}], \forall i$. {The energy output of storage in the $i^{\text{th}}$ instant is given by $s_i = \frac{[x_i]^+}{\eta_{ch}} - \eta_{dis} [x_i]^-$, where $[x]^+$=$\max(0, x)$ and $[x]^-$=$-\min(0, x)$.} 
The ramping constraint induce limits on $s_i$ given by
\begin{equation}
s_i \in [\delta_{\min}h\eta_{dis}, \delta_{\max}h/\eta_{ch}], \quad \forall i. \vspace{-3pt}
\label{constraintramp}
\end{equation}
The energy stored in the battery is denoted as $b_i$, defined as $b_i = b_{i-1}+ x_i$. Battery capacity constraint is given as
\begin{equation}
b_i \in [b_{\min}, b_{\max}], \quad \forall i. \vspace{-3pt}
\label{constraintcapacity}
\end{equation}
where $b_{\min}$ and $b_{\max}$ are minimum and maximum permissible battery charge levels respectively. We use xC-yC notation to represent the relationship between ramp rate and battery capacity. xC-yC implies battery takes 1/x hours to charge and 1/y hours to discharge completely.

\subsection{Cost of the battery with inverter}
The cost of inverter depends on the rated apparent power output capability. Based on cost of 6 kW SMA Sunny Boy 2.0, the cost per kW of inverter is 100 euros/kWh \cite{smacost}.
~The cost of Li-Ion battery is assumed to have two components: (a) capacity cost and (b) ramping capability cost, similar to \cite{batcostweb}. The cost of battery (with inverter) for per kWh in euros is given as 
\begin{equation}
\text{Battery Cost} = 300 + {0.25}~{\max(x,y)} 100 ~~ \text{per kWh}, \vspace{-4pt}
\end{equation} 
where $x,y$ denotes the charging and discharging rates as described in the battery model xC-yC.

\subsection{Co-optimization formulation}
\begin{figure}[!htbp]
	\center
	\includegraphics[width=3.2in]{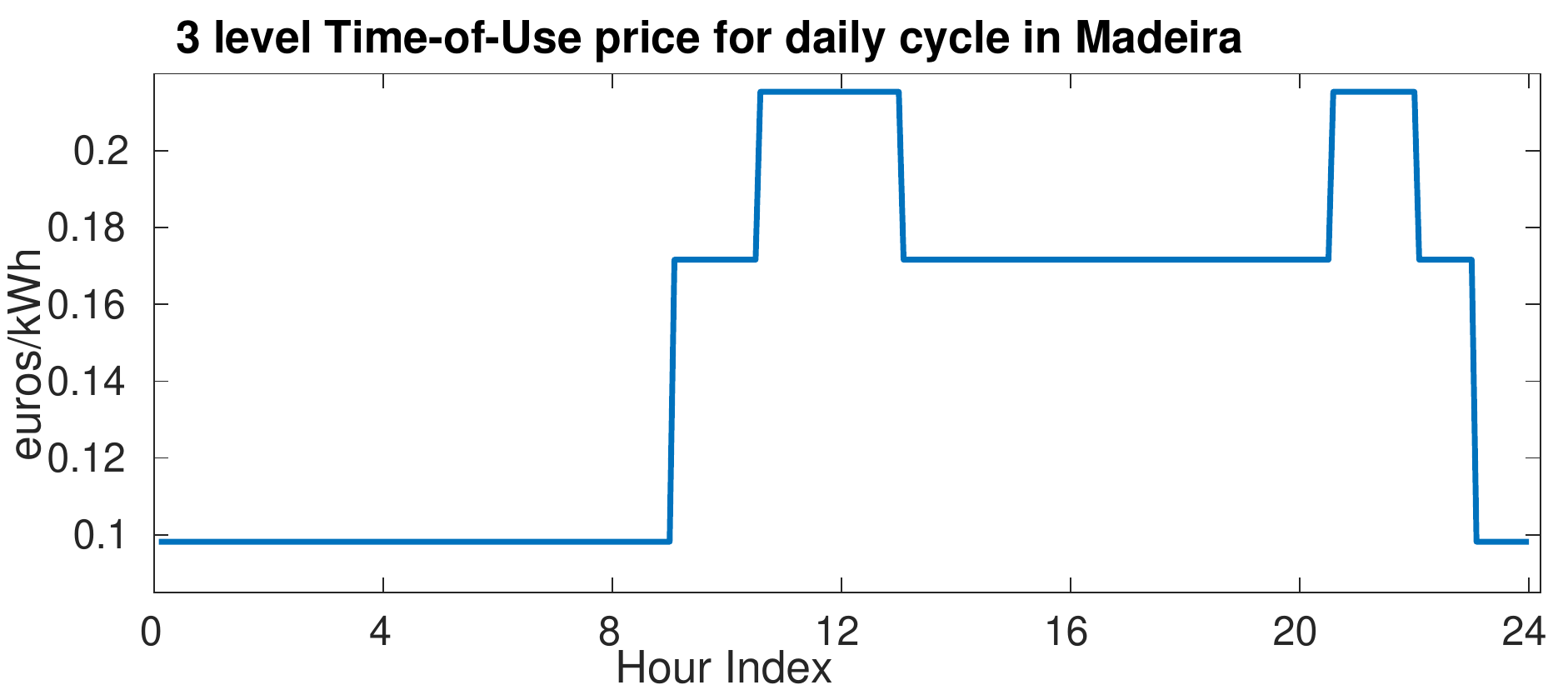}
	\vspace{-10pt}
	\caption{{Time-of-use (ToU) electricity prices}}
	\label{fig:tou}
\end{figure}

The power network in Madeira impose time-of-use (ToU) electricity prices for consumption, applies peak power restriction and imposes zero feed-in-tariff. The ToU price considered is shown in Fig.~\ref{fig:tou}. The PPC contracts have 8 levels: 3.45, 4.6, 5.75, 6.9, 10.35, 13.8, 17.25 and 20.7 kVA for LV prosumers with per day cost as 0.1643, 0.2132, 0.2590, 0.3080, 0.4532, 0.5981, 0.7436 and 0.8892 respectively. The detailed consumer contracts in Madeira are summarized in \cite{hashmi2019energy}.
The co-optimization formulation is developed in \cite{hashmi2019energy} and given as
\begin{mini!}[1]
	{s_i}{ \sum_{i=1}^N {p}_{\text{elec}}(i)\theta_i(s_i) h}{}{(P_{opt})~~}\vspace{-4pt}
	\label{eq:optprob1}
	\addConstraint{\text{Ramping constraint,~} }{Eq.~\ref{constraintramp}}{ }\vspace{-4pt}
	\label{eq4b}
	\addConstraint{\text{Capacity constraint,~}}{Eq.~\ref{constraintcapacity}}\vspace{-4pt}
	\label{eq4c}
	\addConstraint{\text{Self-sufficiency,}~~\theta_i(s_i) }{\geq 0}\vspace{-4pt}
	\label{eq4d}
	\addConstraint{\text{Arbitrage,}~~\theta_i(s_i) }{\geq [z_i + s_i]}\vspace{-4pt}
	\label{eq4e}
	\addConstraint{\text{Peak shaving,}~~[z_i + s_i]/h }{\leq P_{\max}^{set}}\vspace{-4pt}
	\label{eq4f}
\end{mini!}

where $\theta_i(s_i)=\max(0, z_i + s_i)$.
The peak power threshold, $P_{\max}^{set}$, is selected close to the power level ($P_{\max} + \delta_{\min}$), subject to $P_{\max}^{set} \geq (P_{\max} + \delta_{\min})$. $P_{\max}^{set}$ is selected by the electricity consumer as a PPC contract with the utility in Madeira. 
Note that this formulation prioritizes self-consumption over arbitrage.

\subsection{Co-optimization with control of cycles}
Note that the formulations discussed previously do not consider battery life that is affected by charge-discharge cycles. Battery manufacturers measure the life of a battery using two indices: cycle life and calendar life. Cycle life denotes the number of cycles a battery can operate at a certain depth-of-discharge before reaching its end-of-life or EoL (EoL is 
reached when battery capacity reduces to 
80\% of its initial rated capacity in Wh). Similarly, calendar life denotes the maximum probable age that the battery can be operational before reaching EoL. Following our prior work \cite{hashmi2018long}, we define a friction function for the active power to model the degradation due to cycles of operation as
$
P_{\text{fric}}^i = \frac{[P_B^i]^+}{\eta_{\text{fric}}} - {[P_B^i]^-\eta_{\text{fric}}}.
$
In the original formulation ($P_{opt}$) the constraint Eq.~\ref{eq4e} is modified as $\theta_i {\geq [z_i + P_{\text{fric}}^i]}$. 
The friction coefficient takes a value from 1 to 0. 
$\eta_{\text{fric}}$ needs to be tuned so as the operational life is increased by matching calendar and cycle degradation \cite{hashmi2018long}.
If the battery is not over operating then $\eta_{\text{fric}}$ is set to 1. For cases where the battery is over-performing, the low returning transactions is eliminated by decreasing value of $\eta_{\text{fric}}$. 

\section{Energy storage profitability}
We propose a mechanism for ensuring storage profitability considering battery cycle and calendar life, thus in effect considering battery degradation. 
Algorithm \texttt{BatteryProfitability} provides the steps to identify storage profitability. For more than one battery is profitable, in such a case it will depend on the prosumer on whether they prioritize payback period or the returns per cycle.

	\begin{algorithm}
		\small{\textbf{Inputs}: {$\eta_{\text{ch}}, \eta_{\text{dis}}, \delta_{\max}, \delta_{\min}, b_{\max}, b_{\min}$}, $b_0$},  {$h, N, T,i=0 $}\\
		\begin{algorithmic}[1]
			\State Calculate total storage profit, $G_T$, by solving $P_{opt}$ over the instants, $N_{total}$, based on load, price and renewable generation,
			\State Calculate equivalent 100\% DoD cycles denoted as $N_{cyc}^{100}$ using \texttt{DoDofVector} proposed in \cite{hashmi2018long},
			\State Calculate $G_{cyc} = G_T/(N_{cyc}^{100}~~ b_{\text{rated}})$,
			\State Calculate $P_{cyc} = G_{cyc} - C_{cyc}$, where $C_{cyc}$ is the per cycle (100\%) cost of the battery,
			\State Calculate the Expected Payback Period (ExPB) in years $= (B_{cost}N_{total})/(G_T N_{year})$, where $B_{cost}$ denotes the battery cost and $N_{year}$ denotes number of samples in a year,
			\State Battery is profitable if $P_{cyc} > 0$ and ExPB $<$ Calendar life of the battery.
		\end{algorithmic}
		\caption{\texttt{BatteryProfitability}}\label{alg:profitability}
	\end{algorithm}

\vspace{-10pt}
\section{Numerical Results}
For the numerical simulations we use the battery parameters listed in Table~\ref{parameters}. The characteristics of the four prosumers are described in Figure~\ref{fig:upac_pvload} and Table~\ref{tab:upac_details}. Simulations are performed for the month of June 2019.

The performance indices used for evaluation are:\\
$\quad \bullet$ \textit{Arbitrage and self-sufficiency gains} ($G_{arb}$);\\
$\quad \bullet$ \textit{Peak shaving gains} ($G_{\text{PD}}$): difference between nominal PPC and the new PPC contract after adding storage;\\
$\quad \bullet$ \textit{Total gains} ($G_{\text{T}}$): is the sum of $G_{arb}$ and $G_{\text{PD}}$, \\
$\quad \bullet$ \textit{Gains per cycle} ($G_{cyc}$): In prior work \cite{hashmi2018long} we develop a mechanism to measure the number of cycles of operation based on DoD of energy storage operational cycles. \\
$\quad \bullet$ \textit{Profit per cycle} ($P_{cyc}$): is the profit battery owner makes per cycle compared to the per cycle cost of the battery ($C_{cyc}$); $P_{cyc} = G_{cyc} - C_{cyc}$;\\
$\quad \bullet$ \textit{Expected payback period} (ExPB): is the linear extrapolation of the payback period compared to $G_{\text{T}}$. ExPB $=C_{bat}/ G_{\text{T}}$\\
$\quad \bullet$ \textit{Self-sufficiency} (SS): calculated using the total energy consumed, PV generation and storage output,\\
$\quad \bullet$ \textit{Wasted energy}: The surplus production is wasted energy is measured in kWh.


We consider the battery cycle life equals 4000 cycles at 100\% DoD. Using this description the euros per cycle per rated battery capacity for different ramping batteries are listed in Table~\ref{returnscomparison}. Considering the calendar life equals 7 years, the battery should perform $\approx$47.6 cycles per month in order to last 7 years and make more than the values listen in Table~\ref{returnscomparison} to be profitable.

\begin{table}[!htbp]
	\scriptsize
		\caption {Battery Parameters}
		\label{parameters}
		\vspace{-7pt}
		\begin{center}
			\begin{tabular}{| c | c|}
				\hline
				$b_{\min}$, $b_{\max}$, $b_{0}$ & 10\%, 100\%, 50\% of $b_{\text{rated}}$\\
				$b_{\text{rated}}$ & 1, 2, 5 kWh\\
				\hline
				$\eta_{\text{ch}}=\eta_{\text{dis}}$ & 0.95\\
				\hline
				$\delta_{\max} = - \delta_{\min}$ & 0.25 $b_{\text{rated}}$ W for 0.25C-0.25C, \\
				& $b_{\text{rated}}$ W for 1C-1C, ~~ 2$b_{\text{rated}}$ W for 2C-2C\\
				\hline
			\end{tabular}
			\hfill\
		\end{center}
	\end{table}
\begin{table}[!htbp]
	\scriptsize
	\caption {Storage profitability with different ramping per kWh}
	\label{returnscomparison}
	\vspace{-7pt}
	\begin{center}
		\begin{tabular}{| c| c| c|c| c|}
			\hline
			Battery  & Inverter  & Battery  & Battery Cost&  euros/cycle/$b_{\text{rated}}$\\
			Model &  Cost/kWh & Cost/kWh & ($C_{\text{bat}}$/kWh) & ($C_{\text{cyc}}$)\\
			\hline
			0.25C-0.25C 	& 25 & 400 & 425&  0.1062\\
			1C-1C	& 100 & 600 & 700 & 0.1750 \\
			2C-2C	& 200 & 700 & 900 & 0.2250 \\
			\hline
		\end{tabular}
		\hfill\
	\end{center}
\end{table}

\subsection{Co-Optimization and storage profitability}
The co-optimization results for each prosumer are presented in Tables \ref{tab:upac2_5min}, \ref{tab:upac3_5min}, \ref{tab:upac8_5min} and \ref{tab:upac12_5min}. 
{U2, U3, U8 and U12 belong to categories C1, C2, C3 and C4, respectively.}

\begin{table}[ht]
\scriptsize
	\caption {(C1) U2: PV generation slightly more than inelastic load}
	\label{tab:upac2_5min}
	\begin{center}
	\vspace{-14pt}
		\begin{tabular}{|c||c|c|c|c|c|c|c|}
			\hline
			\multirow{2}{*}{Case} &  $G_{\text{PD}}$ & $G_{T}$ & $P_{\text{cyc}}$ & Cycles & ExPB &SS & Waste  \\ 
				        & Euro & Euro &  &  &(years)  & (\%) & (kWh) \\
			\hline
			Load + PV  & - & - & - & - & - &26.82 & 57.4 \\
			\hline
			\multicolumn{8}{|c|}{\textbf{1 kWh }Battery} \\
			\hline
            \hl{\textbf{0.25C-0.25C}} &  0 & 10.13 & {\textbf{0.1675}} &  37.01  &  \textbf{3.50} & 31.91 & 30.12 \\
            \hl{\textbf{1C-1C} }		&  1.39 & 13.79 & {\textbf{0.0867}}  & \textcolor{red}{52.70} &  \textbf{4.23}   & 33.56 & 18.59 \\
            \hl{\textbf{2C-2C }}		& 2.85 & 15.48 & {\textbf{0.0536}} &  \textcolor{red}{55.56} &  \textbf{4.84}   & 33.69 & 15.39 \\
			\hline
			\multicolumn{8}{|c|}{\textbf{2 kWh} Battery}\\
			\hline
			\hl{\textbf{0.25C-0.25C}} & 1.39 & 15.83 &\textbf{ 0.0798} & 42.53  &    \textbf{4.47}   & 34.96 & 12.97 \\
            \textbf{1C-1C} 		& 2.85 & 19.26 & {\textbf{0.0355}}  &   45.75 &  \textbf{6.06}    & 35.85 & 5.15 \\
            2C-2C 		& 2.85 & 19.33 & -0.016 &  46.24&      7.76   & 35.88 & 3.96 \\
			\hline
			\multicolumn{8}{|c|}{5 kWh Battery} \\
			\hline
			0.25C-0.25C &  1.39 & 22.46 & \textbf{0.0288}  & 33.27     &   7.88    & 36.38 & 0.39 \\
            1C-1C 		&  2.85 & 24.67 & -0.0295 &   33.91 &  11.82    &      36.29  & 0.19 \\
            2C-2C 		&  2.85 & 24.67 & -0.0795 & 33.91   &      15.20  &   36.29 & 0.18 \\
			\hline
		\end{tabular}
		\hfill\
	\end{center}
\end{table}

%

The best-suited battery is selected based on: (P1) profit per cycle per unit capacity, (P2) cycles of operation and (P3) expected payback period. 

The best-suited battery for U2, U3, U8 and U12 are marked in bold in 
Tables \ref{tab:upac2_5min}, \ref{tab:upac3_5min}, \ref{tab:upac8_5min}, and \ref{tab:upac12_5min} respectively. 
For battery installation to make sense the value of $P_{\text{cyc}}$ should be positive and ExPB should be lower than 7 years (i.e. the calendar life of battery). For prosumer U2 5 batteries satisfy the feasibility criteria. However, the best-suited battery will vary with preference. For instance if payback period is priority the U2 should select 1 kWh 0.25C-0.25C battery. 
Note that the marginal value of storage decreases with increase in ramping capability and charge capacity. Thus if $P_{cyc}$ is used to select the battery then slowest ramping battery with smallest capacity will be selected.

%

The self-sufficiency increases with increase in battery capacity and ramping capability thus the wasted excess generation decreases.  

\begin{table}[ht]
\scriptsize
	\caption {(C2) U3: Active load management to reduce excess generation}
	\label{tab:upac3_5min}
	\begin{center}
	\vspace{-7pt}
		\begin{tabular}{|c||c|c|c|c|c|c|c|}
			\hline
			\multirow{2}{*}{Case} &  $G_{\text{PD}}$ & $G_{T}$ & $P_{\text{cyc}}$ & Cycles & ExPB &SS & Waste  \\ 
				        & Euro & Euro &  &  &(years)  & (\%) & (kWh) \\
			\hline
			Load + PV  & - & - & - & - & - &32.71 & 2.72 \\
			\hline
			\multicolumn{8}{|c|}{\textbf{1 kWh} Battery} \\
			\hline
            \hl{\textbf{0.25C-0.25C}} & 0 & 7.57 &\textbf{0.1715 }&  27.26  & \textbf{4.68}     &32.73 & 0.01 \\
          { 1C-1C} 		&  0 & 8.13 & {\textbf{0.0197}}&   41.76 &  7.18  & 32.53 & 0 \\
            2C-2C 		&  0 & 8.14 & -0.0318 &   42.13   &    9.21  & 32.52 & 0 \\
			\hline
			\multicolumn{8}{|c|}{2 kWh Battery}\\
			\hline
			0.25C-0.25C & 0 & 9.74 & \textbf{0.0791}& 26.28 &   7.27   & 32.09 & 0 \\
            1C-1C 		&  0 & 10.07 & \textbf{0.0065} & 27.74 &  11.59   & 32.04 & 0 \\
            2C-2C 		& 0 & 10.07 & -0.0436 &   27.76  &   14.90  & 32.04 & 0 \\
			\hline
			\multicolumn{8}{|c|}{5 kWh Battery} \\
			\hline
            0.25C-0.25C &  0 & 13.92 & 0.0189 &   22.25&    12.72   & 30.66 & 0 \\
            1C-1C 		& 0 & 14.02 & -0.05 & 22.43 &      20.80  & 30.64 & 0 \\
            2C-2C 		&  0 & 14.02 & -0.1 &   22.43 &    26.75    & 30.64 & 0 \\
			\hline
		\end{tabular}
		\hfill\
	\end{center}
\end{table}

\begin{table}[ht]
\scriptsize
	\caption {(C3) U8: Comparable generation and load with average load higher then C1,C2 and C4}
	\label{tab:upac8_5min}
	\begin{center}
	\vspace{-14pt}
		\begin{tabular}{|c||c|c|c|c|c|c|c|}
			\hline
			\multirow{2}{*}{Case} &  $G_{\text{PD}}$ & $G_{T}$ & $P_{\text{cyc}}$ & Cycles & ExPB &SS & Waste  \\ 
				        & Euro & Euro &  &  &(years)  & (\%) & (kWh) \\
			\hline
			Load +PV  & - & - & - & - & - & 28.02 & 58.14 \\
			\hline
			\multicolumn{8}{|c|}{\textbf{1 kWh} Battery} \\
			\hline
            \hl{\textbf{0.25C-0.25C }}&  0 & 35.62 & \textbf{0.8589} &   36.91 &  \textbf{0.99}    & 28.71 & 40.68 \\
            \hl{\textbf{1C-1C }}		&  0 & 39.42 & \textbf{0.4543} &   \textcolor{red}{62.64}  &  \textbf{1.48}    & 29.42 & 22.69 \\
           \hl{\textbf{ 2C-2C }}		& 0 & 40.02 & \textbf{0.3765} &  \textcolor{red}{66.53}   &   \textbf{1.87 }   & 29.57 & 17.88 \\
			\hline
			\multicolumn{8}{|c|}{\textbf{2 kWh} Battery}\\
			\hline
		\hl{\textbf{	0.25C-0.25C}} & 0 & 40.04 & \textbf{0.2941} &  50.01   &   \textbf{1.77 }  & 29.21 & 27.27 \\
            \hl{\textbf{1C-1C}} 		&  0 & 44.36 & \textbf{0.1842}  &  \textcolor{red}{61.75}   &   \textbf{2.63}  & 29.79 & 10.73 \\
            \hl{\textbf{2C-2C }}		&  4.31 & 48.70 & \textbf{0.1774}  & \textcolor{red}{60.51}   &    \textbf{3.08}   & 29.79 & 9.09 \\
			\hline
			\multicolumn{8}{|c|}{\textbf{5 kWh} Battery} \\
			\hline
            \hl{\textbf{0.25C-0.25C}} & 0 & 50.44 & {\textbf{0.1117}}  &    46.30    & \textbf{3.51}   & 29.90 & 4.80 \\
            \hl{\textbf{1C-1C }}		& 4.31 & {58.64} & {\textbf{0.1106}} &  41.06    &  \textbf{4.97}   & 29.80 & 1.54 \\
            \hl{\textbf{2C-2C} }		&  4.31 & 58.65 & \textbf{0.0598} &  41.18    &   \textbf{6.39}   &     29.80 & 1.36 \\
			\hline
		\end{tabular}
		\hfill\
	\end{center}
\end{table}
\begin{table}[ht]
\scriptsize
	\caption {(C4) U12: PV generation significantly more than inelastic load}
	\label{tab:upac12_5min}
	\begin{center}
	\vspace{-14pt}
		\begin{tabular}{|c||c|c|c|c|c|c|c|}
			\hline
			\multirow{2}{*}{Case} &  $G_{\text{PD}}$ & $G_{T}$ & $P_{\text{cyc}}$ & Cycles & ExPB &SS & Waste  \\ 
				        & Euro & Euro &  &  &(years)  & (\%) & (kWh) \\
			\hline
			Load + PV  & - & - & - & - & - & 47.84 & 158.1 \\
			\hline
			\multicolumn{8}{|c|}{\textbf{1 kWh }Battery} \\
			\hline
            \hl{\textbf{0.25C-0.25C}} &  0 & 8.57 & \textbf{0.1922} & 28.69  &  \textbf{4.13 }  & 56.91 & 126.13 \\
            \hl{\textbf{1C-1C }}		& 0 & 9.49 & \textbf{0.1107} &   33.22   &   \textbf{6.15}   & 57.72 & 115.66 \\
            2C-2C 		&  0 & 9.56 & \textbf{0.0739} & 35.55   &  7.85     & 57.84 & 106.63 \\
			\hline
			\multicolumn{8}{|c|}{\textbf{2 kWh} Battery}\\
			\hline
            \hl{\textbf{0.25C-0.25C}} &  0 & 13.41 & {\textbf{0.0566}}  &   41.19 &    \textbf{5.28}   &     65.00 & 97.50 \\
            1C-1C 		& 0 & 13.84 & \textbf{0.0214} &  35.23    &   8.43   & 65.77 & 81.94 \\
            2C-2C 		&  0 & 13.86 & -0.0116 &  32.47    &   10.82  & 65.80 & 70.04 \\
			\hline
			\multicolumn{8}{|c|}{5 kWh Battery} \\
			\hline
            0.25C-0.25C & 0 & 18.28 & -0.007 &  36.85    &   9.69  & 74.90 & 53.58 \\
            1C-1C 		&  0 & 18.71 & -0.0758 &      37.72   &   15.59   & 75.66 & 31.16 \\
            2C-2C 		&  0 & 18.70 & -0.1257  &  37.66    &  20.05  & 75.65 & 27.41 \\
			\hline
		\end{tabular}
		\hfill\
	\end{center}
\end{table}
%

\subsection{Friction Coefficient}
The friction coefficient can help in reducing the cycles of operation for cases where the battery is performing more number of cycles compared to the value which matches the calendar life degradation. 
For equalizing cycle and calendar life degradation of the battery, it should perform $\approx 47.6$ cycles
in 1 month (considering 4000 as cycle life at 100\% DoD and 7 years as calendar life).
From Tables \ref{tab:upac2_5min}, \ref{tab:upac3_5min}, \ref{tab:upac8_5min}, and \ref{tab:upac12_5min} it is clear that inclusion of friction coefficient for limiting cycles of operation is only relevant for 1 and 2 kWh batteries for U8 and 1 kWh battery for U2, as for these case the battery is operating more number of cycles than the optimal value of 47.6 cycles in a month (marked in red). 
Table \ref{tab:friction_coefficient} compares the tuned friction coefficients for batteries which performed more cycles than expected. 
Note that friction coefficient can only reduce cycles of operation, therefore it is only applicable when the battery is over-performing.
Inclusion of $\eta_{\text{fric}}$ for over-performing batteries increases $P_{cyc}$, however, also increases ExPB.

\begin{table}[!tbph]
	\scriptsize
	\caption {Friction Coefficient.}
	\label{tab:friction_coefficient}
	\vspace{-7pt}
	\begin{center}
		\begin{tabular}{|c||c|c|c|c|c|c| c|}
			\hline
			Battery &$\eta_{\text{fric}}$ & Profit & $P_{cyc}$ & SS & Waste & Cycles  & ExPB \\ 
			Model& & euros &  & (\%) & (kWh) &  & yrs. \\
			\hline
			\multicolumn{8}{|c|}{For prosumer U2} \\
			\hline
            \textbf{1kWh,1C-1C} & 0.796 & 12.85 & 0.0962 & 32.45 & 25.70 & 47.38 & 4.54 \\
            \textbf{1kWh,2C-2C} & 0.797 & 14.45 & 0.0796 & 32.53 & 24.99 & 47.44 & 5.19 \\
			\hline
			\multicolumn{8}{|c|}{For prosumer U8}\\
			\hline
            \textbf{1kWh,1C-1C} & 0.796 & 38.17 & 0.6805 & 29.33 & 26.82 & 44.62 & 1.53 \\
            \textbf{1kWh,2C-2C} & 0.796 & 38.50 & 0.5829 & 29.37 & 25.16 & 47.65 & 1.95 \\
			\hline
			\textbf{2kWh,1C-1C} & 0.9398 &43.98 & 0.3143 & 29.67 & 13.71 & 52.29 & 1.61 \\
			\textbf{2kWh,1C-1C} & 0.9397 & 43.47 & 0.3286 & 29.73 & 14.34 & 43.17 & 2.68\\
            \textbf{2kWh,2C-2C} & 0.9397 & 47.85 & 0.3228 & 29.76 & 12.91 & 43.68 & 3.13 \\
			\hline
		\end{tabular}
		\hfill\
	\end{center}
\end{table}
\section{Conclusion}
The profitability of investment in storage is a crucial factor for deciding future investment in batteries. In this paper, we co-optimize storage in the context of power network norms in Madeira. The batteries are used for performing energy arbitrage, increasing self-sufficiency and peak demand reduction. 

The proposed formulation prioritizes self-sufficiency over energy arbitrage, thus minimizing the excess production at the prosumer end.
Four types of prosumers are selected based on the different relationship between inelastic load and renewable generation to identify the value of storage. 

It is observed that the marginal value of installing battery decreases with storage size and ramping capability. 
We observe that value of storage for an average net-load comparable to storage ramping rate leads to profits several folds higher than for otherwise.
Faster ramping batteries perform much more number of cycles which deteriorates the profit made per 100\% DoD cycles per unit of storage capacity making such batteries financially unviable. 

We propose the inclusion of friction coefficient which reduces the cycles of operation by eliminating storage operations for low returning transactions thus increasing the profit per cycle per unit of storage capacity and makes such batteries more profitable, however, it increases the payback period.

\section*{Acknowledgment}
{Funding from grants: ANR under ANR-16-CE05-0008, EU H2020 under GA 731249, and FCT under UID/EEA/50009/2019 are gratefully acknowledged.}



\bibliographystyle{IEEEtran}
\bibliography{IEEEabrv,biblio}

\begin{thebibliography}{10}
\providecommand{\url}[1]{#1}
\csname url@samestyle\endcsname
\providecommand{\newblock}{\relax}
\providecommand{\bibinfo}[2]{#2}
\providecommand{\BIBentrySTDinterwordspacing}{\spaceskip=0pt\relax}
\providecommand{\BIBentryALTinterwordstretchfactor}{4}
\providecommand{\BIBentryALTinterwordspacing}{\spaceskip=\fontdimen2\font plus
\BIBentryALTinterwordstretchfactor\fontdimen3\font minus
  \fontdimen4\font\relax}
\providecommand{\BIBforeignlanguage}[2]{{%
\expandafter\ifx\csname l@#1\endcsname\relax
\typeout{** WARNING: IEEEtran.bst: No hyphenation pattern has been}%
\typeout{** loaded for the language `#1'. Using the pattern for}%
\typeout{** the default language instead.}%
\else
\language=\csname l@#1\endcsname
\fi
#2}}
\providecommand{\BIBdecl}{\relax}
\BIBdecl

\bibitem{hashmi2018effect}
M.~U. Hashmi, D.~Muthirayan, and A.~Bu{\v{s}}i{\'c}, ``Effect of real-time
  electricity pricing on ancillary service requirements,'' in \emph{9th
  e-Energy}.\hskip 1em plus 0.5em minus 0.4em\relax ACM, 2018, pp. 550--555.

\bibitem{shi2018using}
Y.~Shi, B.~Xu, D.~Wang, and B.~Zhang, ``Using battery storage for peak shaving
  and frequency regulation: Joint optimization for superlinear gains,''
  \emph{IEEE Transactions on Power Systems}, vol.~33, no.~3, pp. 2882--2894,
  2018.

\bibitem{hashmi2019energy}
M.~U. Hashmi, L.~Pereira, and A.~Bu{\v{s}}i{\'c}, ``Energy storage in madeira,
  portugal: Co-optimizing for arbitrage, self-sufficiency, peak shaving and
  energy backup,'' \emph{arXiv preprint arXiv:1904.00463}, 2019.

\bibitem{Kied1910:Sensitivity}
D.~Kiedanski, M.~U. Hashmi, A.~Busic, and D.~Kofman, ``Sensitivity to forecast
  errors in energy storage arbitrage for residential consumers,'' in
  \emph{SmartGridComm}, Beijing, P.R. China, Oct. 2019.

\bibitem{hashmi2017optimal}
M.~U. Hashmi, A.~Mukhopadhyay, A.~Bu{\v{s}}i{\'c}, and J.~Elias, ``Optimal
  control of storage under time varying electricity prices,'' in
  \emph{SmartGridComm}.\hskip 1em plus 0.5em minus 0.4em\relax IEEE, 2017, pp.
  134--140.

\bibitem{Hash1910:Optimal}
M.~U. Hashmi, A.~Mukhopadhyay, A.~Busic, J.~Elias, and D.~Kiedanski, ``Optimal
  storage arbitrage under net metering using linear programming,'' in
  \emph{SmartGridComm}, Beijing, P.R. China, Oct. 2019.

\bibitem{oudalov2007sizing}
A.~Oudalov, R.~Cherkaoui, and A.~Beguin, ``Sizing and optimal operation of
  battery energy storage system for peak shaving application,'' in \emph{Power
  Tech, 2007 IEEE Lausanne}.\hskip 1em plus 0.5em minus 0.4em\relax IEEE, 2007,
  pp. 621--625.

\bibitem{shen2014optimization}
J.~Shen, S.~Dusmez, and A.~Khaligh, ``Optimization of sizing and battery cycle
  life in battery/ultracapacitor hybrid energy storage systems for electric
  vehicle applications,'' \emph{IEEE Trans. on industrial informatics},
  vol.~10, no.~4, pp. 2112--2121, 2014.

\bibitem{bahramirad2012reliability}
S.~Bahramirad, W.~Reder, and A.~Khodaei, ``Reliability-constrained optimal
  sizing of energy storage system in a microgrid,'' \emph{IEEE Trans. on Smart
  Grid}, vol.~3, no.~4, pp. 2056--2062, 2012.

\bibitem{chen2011sizing}
S.~Chen, H.~B. Gooi, and M.~Wang, ``Sizing of energy storage for microgrids,''
  \emph{IEEE Trans. on Smart Grid}, vol.~3, no.~1, pp. 142--151, 2011.

\bibitem{brekken2010optimal}
T.~K. Brekken, A.~Yokochi, A.~Von~Jouanne, Z.~Z. Yen, H.~M. Hapke, and D.~A.
  Halamay, ``Optimal energy storage sizing and control for wind power
  applications,'' \emph{IEEE Transactions on Sustainable Energy}, vol.~2,
  no.~1, pp. 69--77, 2010.

\bibitem{megel2014scheduling}
O.~M{\'e}gel, J.~L. Mathieu, and G.~Andersson, ``Scheduling distributed energy
  storage units to provide multiple services,'' in \emph{PSCC}.\hskip 1em plus
  0.5em minus 0.4em\relax IEEE, 2014.

\bibitem{walawalkar2007economics}
R.~Walawalkar, J.~Apt, and R.~Mancini, ``Economics of electric energy storage
  for energy arbitrage and regulation in new york,'' \emph{Energy Policy},
  vol.~35, no.~4, pp. 2558--2568, 2007.

\bibitem{anderson2017co}
K.~Anderson and A.~El~Gamal, ``Co-optimizing the value of storage in energy and
  regulation service markets,'' \emph{Energy Systems}, vol.~8, no.~2, pp.
  369--387, 2017.

\bibitem{cheng2016co}
B.~Cheng and W.~Powell, ``Co-optimizing battery storage for the frequency
  regulation and energy arbitrage using multi-scale dynamic programming,''
  \emph{IEEE Trans. on Smart Grid}, 2016.

\bibitem{white2011using}
C.~D. White and K.~M. Zhang, ``Using vehicle-to-grid technology for frequency
  regulation and peak-load reduction,'' \emph{Journal of Power Sources}, vol.
  196, no.~8, pp. 3972--3980, 2011.

\bibitem{dvorkin2016ensuring}
Y.~Dvorkin, R.~Fernandez-Blanco, D.~S. Kirschen, H.~Pand{\v{z}}i{\'c}, J.-P.
  Watson, and C.~A. Silva-Monroy, ``Ensuring profitability of energy storage,''
  \emph{IEEE Transactions on Power Systems}, vol.~32, no.~1, pp. 611--623,
  2016.

\bibitem{hashmi2019arbitrage}
M.~U. Hashmi, D.~Deka, A.~Busic, L.~Pereira, and S.~Backhaus, ``Arbitrage with
  power factor correction using energy storage,'' \emph{arXiv preprint
  arXiv:1903.06132}, 2019.

\bibitem{hashmi2018limiting}
M.~U. Hashmi and A.~Busic, ``Limiting energy storage cycles of operation,'' in
  \emph{GreenTech}.\hskip 1em plus 0.5em minus 0.4em\relax IEEE, 2018, pp.
  71--74.

\bibitem{hashmi2018long}
M.~U. Hashmi, W.~Labidi, A.~Bu{\v{s}}i{\'c}, S.-E. Elayoubi, and T.~Chahed,
  ``Long-term revenue estimation for battery performing arbitrage and ancillary
  services,'' in \emph{SmartGridComm}.\hskip 1em plus 0.5em minus 0.4em\relax
  IEEE, 2018.

\bibitem{mackley2015technical}
R.~D. Mackley, D.~M. Anderson, J.~N. Thomle, and C.~E. Strickland, ``Technical
  and economic assessment of solar photovoltaic for groundwater extraction on
  the hanford site,'' PNNL, Richland, WA (United States), Tech. Rep., 2015.

\bibitem{bnefbattery}
\BIBentryALTinterwordspacing
``A behind the scenes take on lithium-ion battery prices,'' BloombergNEF
  article, 2019. [Online]. Available: \url{https://tinyurl.com/y3xbu8xt}
\BIBentrySTDinterwordspacing

\bibitem{pereira_value_2019}
L.~Pereira and J.~Cavaleiro, ``On the {Value} {Proposition} of {Battery}
  {Energy} {Storage} in {Self}-{Consumption} {Only} {Scenarios}: {A}
  {Case}-{Study} in {Madeira} {Island},'' in \emph{{IECON} 2019 - 45th {Annual}
  {Conference} of the {IEEE} {Industrial} {Electronics} {Society}}, Lisbon,
  Portugal, Oct. 2019.

\bibitem{prsma_data_2018}
{Prsma}, {M-ITI}, {EEM}, and {ACIF-CCIM}, ``\BIBforeignlanguage{en}{Data
  {Collection}, {Modelling}, {Simulation} and {Decision}},'' European
  Commission, Funchal, Portugal, Technical report 4.3, Jun. 2018.

\bibitem{prsma_detailed_2018}
{Prsma}, {EEM}, {M-ITI}, {ACIF-CCIM}, {LIBAL}, and {DTU},
  ``\BIBforeignlanguage{en}{Detailed {Plan} of {Action} for the {DSM}
  {Demonstrator}},'' European Commission, Funchal, Portugal, Technical report
  4.6, Jul. 2018.

\bibitem{smacost}
\BIBentryALTinterwordspacing
``Sma sunny boy 2.0 cost,'' 2019. [Online]. Available:
  \url{https://tinyurl.com/yxc8td25}
\BIBentrySTDinterwordspacing

\bibitem{batcostweb}
\BIBentryALTinterwordspacing
C.~Z. Todd~Aquino and C.~Koss, ``Energy storage technology assessment (prepared
  for public service company of new mexico); hdr report no.
  10060535-0zp-c1001,'' 2017. [Online]. Available:
  \url{https://tinyurl.com/y2aeugef}
\BIBentrySTDinterwordspacing

\end{thebibliography}
%



\end{document}